\documentclass[12pt,twoside]{article}
\usepackage{latexsym,amsfonts,amssymb,amsbsy}

\newtheorem{theorem}{THEOREM}
\newtheorem{lemma}{LEMMA}
\newtheorem{proposition}{PROPOSITION}
\newtheorem{corollary}{COROLLARY}
\newcommand{\bthm}{\begin{theorem}\hspace{-2mm}{\bf.\ }}
\newcommand{\ethm}{\end{theorem}}
\newcommand{\blemma}{\begin{lemma}\hspace{-2mm}{\bf.\ }}
\newcommand{\elemma}{\end{lemma}}
\newcommand{\bprop}{\begin{proposition}\hspace{-2mm}{\bf.\ }}
\newcommand{\eprop}{\end{proposition}}
\newcommand{\bcor}{\begin{corollary}\hspace{-2mm}{\bf.\ }}
\newcommand{\ecor}{\end{corollary}}

\def\lover#1#2{\lower #1 pt\hbox{$#2$}}

\begin{document}
\pagestyle{empty}
\vspace* {13mm}
\baselineskip=24pt
\begin{center}
{\bf THE ENERGY OPERATOR FOR A MODEL WITH A MULTIPARAMETRIC
 INFINITE STATISTICS}
\\[11mm]

Stjepan Meljanac $^1$, Ante Perica $^1$ \\
and Dragutin Svrtan $^2$

{\it $^1$ Rudjer Bo\v skovi\'c Institute - Bijeni\v cka c. 54, 10000 Zagreb, Croatia \\[1mm]

$^2$ Dept. of Math., Univ. of Zagreb, Bijeni\v cka c. 30, 10000 Zagreb, Croatia}

\end{center}

\newpage

{\bf Abstract}. In this paper we consider energy operator (a free
Hamiltonian), in the second-quantized approach, for the
multiparameter quon algebras:  $a_{i}a_{j}^{\dagger} -
q_{ij}a_{j}^{\dagger}a_{i} = \delta_{ij},\ i,j\in I$ with
$(q_{ij})_{i,j\in I}$  any hermitian matrix of deformation
parameters. We obtain an elegant formula for normally ordered (sometimes called Wick-ordered) series
expansions of number operators (which determine a free
Hamiltonian).
As a main result (see Theorem 1) we prove that the number
operators are given, with respect to a basis formed by
"generalized Lie elements", by certain normally ordered quadratic
expressions with coefficients given precisely by the entries of
the inverses of Gram matrices of multiparticle weight spaces.
(This settles a conjecture of two of the authors (S.M and A.P),
stated in [8]). These Gram matrices are hermitian generalizations
of the Varchenko's matrices, associated to a quantum (symmetric)
bilinear form of diagonal arrangements of hyperplanes (see [12]).
The solution of the inversion problem of such matrices  in [9]
(Theorem 2.2.17), leads to an effective formula for the number
operators studied in this paper.\\
The one parameter case, in the monomial basis,
was studied by  Zagier [15], Stanciu [11] and M\o ller [6].

\begin{center}
\end{center}

\vskip 0.3cm
\vskip 1.5cm
\newpage
\setcounter{page}{1}
\pagestyle{plain}
\def\leer{\vspace{5mm}}
\setcounter{equation}{0}
\noindent {\bf 1. Introduction.} One-parameter quonic intermediate
statistics [2][3][4], which interpolate between Bose-Einstein and
Fermi-Dirac statistics, are examples of infinite statistics in
which any representation of the symmetric group can occur. These
models offer a possibility of a small violation of the Pauli
exclusion principle, at least in nonrelativistic theory [3][5]. In
a seminal paper [15], Zagier made an explicit computation  of the
Gram determinants of multiparticle weight spaces of the Fock
representation (which for $q\in \langle -1,1\rangle$ proves a
Hilbert space realizability of "q-mutator relations "
$a_{i}a_{j}^{\dagger} - qa_{j}^{\dagger}a_{i} = \delta_{ij},\
i,j\in I $) and begun a study  of particle number operators. A
slight variation of the Zagier's conjecture [15] on the form of a
normally ordered series expansion of the number operators in a
monomial basis is proved subsequently by Stanciu in [11].
Generally, physical observables in the second-quantized approach
are represented in terms of creation and annihilation operators in
the normally ordered form (see M\o ller [6]). Meljanac and Perica
started (in [7], [8]) with an idea to extend the above results to
the multi-parameter case: $a_{i}a_{j}^{\dagger} -
q_{ij}a_{j}^{\dagger}a_{i} = \delta_{ij},\ i,j\in I$, where each
commutation relation has its own deformation parameter $q_{ij}$ (a
complex number) satisfying $q_{ji}=(q_{ij})^{*}$ (where ${}^*$
denotes complex conjugation).

Subsequently, in [9] (see also [10]) two types of results are proved:\\
{\bf Ad.1.} In case of distinct quantum numbers the
multi-parameter Gram determinants (Theorem 1.9.2) are computed by
extending Zagier's method, which in turn gives also a hermitian
analogue of the Varchenko's determinant of the (symmetric) quantum
bilinear form of diagonal arrangements of hyperplanes. From this
explicit computation a Hilbert space realizability follows in case
when all $|q_{ij}| <1$ (cf.\ other methods presented in [16] and [17]).\\
{\bf Ad.2.} Explicit formulas (Theorem 2.2.17) are obtained for
the inverse of the Gram matrices of arbitrary multiparticle weight
spaces, by following ideas of  Bo\v zejko and Speicher (given in
[16]). In particular, a counterexample (when $n=8$) to a
conjecture of Zagier (also stated in [15]), for the form of the
inverse in the one-parameter case, is found. In [9] an appropriate
extension of Zagier's conjecture for the form of the inverse of
multi-parameter Gram matrices is also formulated and proved.

In this paper we study number operators (and hence energy
operator) in the spirit of the second-quantized approach.The
approach is basically algebraic, i.e.\ independent of any
particular representation (see Greenberg [3], M\o ller [6],
Stanciu [11], Meljanac and Perica [8]).

The main result of this paper is the Theorem 1, in which we show
that the coefficients of the normally ordered series expansion of
particle number operators in the Fock representation, in terms of
a basis of ``generalized Lie elements'', are given precisely by
certain inverse matrix entries of the Gram matrices on the
multiparticle weight spaces. This confirms a conjecture of
Meljanac and Perica in [8]. Thus, in conjunction with the results
of [9], one obtains explicit expression for the number operators
in multiparametric quon algebras.

\noindent
 {\bf 2. Multi-parameter quon algebras and Gram matrices.}\\
 Let ${\bf q} = \{ q_{ij}: i,j \in I,({q}_{ij})^{\ast}=q_{ji} \}$ be a
 hermitian family of complex numbers (parameters), where $I$ is a finite
(or infinite) set of indices. Recall that (cf.\ [9]) by a {\em
multiparameter quon algebra} ${\cal A} ={\cal A}^{({\bf q})}$ we
shall mean an associative (complex) algebra generated by $\{
a_{i}, a_{i}^{\dagger}, i \in I \}$ subject to the following
$q_{ij}$- canonical commutation relations:
\[
a_{i}a_{j}^{\dagger} = q_{ij}a_{j}^{\dagger}a_{i} + \delta_{ij},\
\
 \ \forall i,j \in I
\]
The algebra ${\cal A}$ has a cannonical anti-involution
$^{\dagger}:{\cal A}\rightarrow {\cal A}$ (which exchanges $a_{i}$
with $a_{i}^{\dagger}$, reverses products and on the coefficients
acts
by complex conjugation.) \\
 Recall that a Fock representation of ${\cal A}$ is given by a
family of linear operators $a_{i}: {\cal H} \rightarrow {\cal H}$
on a complex Hilbert space ${\cal H}$, $i\in I$, satisfying the
following canonical commutation(or ``$q_{ij}$-mutator'')
relations:
\begin{equation}
a_{i}a_{j}^{\dagger} - q_{ij}a_{j}^{\dagger}a_{i} = \delta_{ij},\ i,j\in I
\end{equation}
\begin{equation}
a_{i}|0\rangle = 0,\ i\in I
\end{equation}
where  $a_{i}^{\dagger}$ denotes
the adjoint of $a_{i}$, and $|0\rangle$ denotes a distinguished (``vacuum'') vector in ${\cal H}$.\\
 Any total order on the indexing  set $I$ induces a total order on the set $I^*$ of all
 sequences (=words) ${\bf i}=i_{1}\dots i_{n}$ over $I$. Then we can consider the Gram matrix \
\begin{equation}
A = (\langle{\bf i}|{\bf j}\rangle)
\end{equation}
\ of all $n$-particle states $|{\bf i}\rangle := a_{i_{1}}
^{\dagger}a_{i_{2}}^{\dagger}\cdots a_{i_{n}}^{\dagger}|0\rangle
(i_{j}\in I, n\geq 0$). Its entries $ \langle {\bf i}|{\bf
j}\rangle$ are  the ``expectation values'' (i.e. overlaps of
$n$-particle states in the second quantized Fock description)
$$\langle 0|a_{i_{n}}\cdots a_{i_{1}}a_{j_{1}}^{\dagger}\cdots a_{j_{m}}^{\dagger}|0\rangle$$
These entries vanish, unless (i) $n=m$ and (ii) $i_{1}\cdots
i_{n}$ and $ j_{1}\cdots j_{m}$ are permutations of the same
weakly increasing sequences ${\nu}=k_{1}\dots k_{n}, k_{1}\leq
\cdots \leq k_{n}, ( k_{j}\in I)$,which we shall call {\it
weights}. Thus the matrix  $A$ is block diagonal (cf.\
[9,Proposition 1.6.1]):
\begin{equation}
A = \oplus_{n\geq 0} \oplus_{k_{1}\leq \cdots \leq k_{n}}A^{k_{1}\dots k_{n}}
\end{equation}
with blocks $A^{\nu}=A^{k_{1}\dots k_{n}}$ indexed by weights. The
size of $A^{\nu}$ is equal to the number of permutations (or
rearrangements) of the multiset $\{k_{1}\leq \cdots \leq k_{n}\}$.

 For ${\nu} = k_{1}<k_{2}<\cdots <k_{n}$ (\ a generic weight ),
$A^{\nu}$ is a matrix of order $n!$ with rows/columns labelled by
rearrangements (of $\nu $)  \mbox{${\bf i}=i_{1}\dots i_{n}=k_{\pi
(1)}\dots k_{\pi (n)}=:\nu.\pi$}, ($\pi \in S_{n} =$ the $n$-th
symmetric group) or simply by permutations $\pi \in S_{n}$. The
entry of $A^{\nu}$ in the row ${\bf i}={\nu}.\pi $ and column
${\bf j}={\nu}.\sigma $ is then given explicitly by the following
formula: \
\begin{equation}
A^{{\nu}}_{{\bf i},{\bf j}} = A^{\nu}(\pi, \sigma ) =
\prod_{(r,s)\in I(\sigma ^{-1}\pi)}q_{k_{\pi (r)}k_{\pi (s)}}
\end{equation}
\noindent
where, for $\pi \in S_{n}$, $I(\pi)$ denotes the set of inversions of
$\pi $:  $I(\pi)=\{(r,s) : 1\leq r<s\leq n, \pi(r)>\pi(s)\}$.
Thus, we can view $A^{\nu}$ as a linear operator on the group algebra
${\bf C}[ S_{n} ]$ = $\{\sum_{\pi \in S_{n}}c_{\pi}\pi $: $c_{\pi}\in
{\bf C}$, $\pi\in S_{n}\}$. \\
 For general weights $\tilde{\nu}=(\tilde{k}_{1}=\cdots
=\tilde{k}_{n_{1}}<\tilde{k}_{n_{1}+1}=\cdots
=\tilde{k}_{n_{1}+n_{2}} <\cdots <\tilde{k}_{n_{1}+\cdots
n_{p-1}+1}=\cdots =\tilde{k}_{n})$, $n_{1}+n_{2}+\cdots +n_{p}=n$,
the matrix $A^{\tilde{\nu}}$ has order equal to $n!/{n_{1}!\cdots
n_{p}!}$ and its rows/columns are labelled by rearrangements ${\bf
i}=i_{1}\dots i_{n} =\tilde{\nu}.\tilde{\pi} $, $\tilde{\pi }\in
H_{\tilde{\nu}} \backslash S_{n}$, where $H_{\tilde{\nu}}={\rm
Stab}_{\tilde{\nu}}=\{ \sigma \in S_{n} | \tilde{\nu}. \sigma
 = \tilde{\nu}\} $ is the (stabilizer) subgroup fixing $\tilde{\nu}$.
The $({\bf i},{\bf j})$-th entry of $A^{\tilde{\nu}}$,
${\bf i}=\tilde{\nu}. \tilde{\pi
 }$, ${\bf j}=\tilde{\nu}.\tilde{\sigma}$, $\tilde{\pi}=H\pi $,
$\tilde{\sigma}=H\sigma $, where $\pi, \sigma $ are unique coset
representatives (of minimal length) of $\tilde{\pi },
\tilde{\sigma }$, is given by
$$A^{\tilde{\nu}}_{{\bf i},{\bf j}}=A^{\tilde{\nu}}(\tilde{\pi },\tilde{\sigma }) \
=\sum_{\tau \in \tilde{\sigma }^{-1}\tilde{\pi }=\sigma ^{-1}H\pi
}\prod_{ (r,s)\in I(\tau )}q_{i_{r}i_{s}}=\sum_{\tau \in
\sigma^{-1}H\pi}\prod _{(r,s)\in I(\tau )}q_{k_{\pi (r)}k_{\pi
(s)}}\ \ \ \ \ \tilde{(5)}$$ \noindent (Note that ($\tilde{5}$)
generalizes (5), because ${\rm Stab}_{\nu}=H_{\nu}=\{1\}$, if
${\nu}$ is generic.) In [9,Subsection 1.7] it is shown that the
operator $A^{\tilde{\nu}}$ can be obtained from $A^{\nu}$
(${\nu}=k_{1}<\dots <k_{n})$ by a {\it reduction procedure} in two steps: first by identifying
indices $k_{1}\mapsto \tilde{k_{1}},\dots, k_{n}\mapsto
\tilde{k_{n}}$ and then restricting this specialized operator
$A^{\nu}|_{{\nu} \mapsto \tilde{\nu}}$ to the invariant subspace
(in ${\bf C}[ S_{n} ]$) spanned by $H_{\tilde{\nu}}$ - invariant
vectors $\overline{\sigma}=\sum_{h\in H_{\tilde{\nu}}}h\sigma
{\in} {\bf C}[ S_{n} ]$. In fact ($\tilde{5}$) can be rewritten as
\begin{equation}
A^{\tilde{\nu}}(\tilde{\pi },\tilde{\sigma }) = \sum_{h\in
H_{\tilde{\nu}}}A^{\nu} (\pi, h\sigma )|_{{\nu} \mapsto
\tilde{\nu}}
\end{equation}
As a consequence we obtain :
 if $A^{\nu}|_{{\nu}\mapsto \tilde{\nu}}$ is invertible, then the
matrix $A^{\tilde{\nu}}$ is invertible too, and a relation
analogous to (6) holds for the inverses. In particular, ${\rm det}
A^{\tilde{\nu}}$ divides ${\rm det} A^{\nu}|_{\nu \mapsto
\tilde{\nu}}$. This shows that in order to study some properties
(e.g. invertibility or positive definiteness)\  it suffices to
consider the generic case (when all the indices $k_{i}$
are distinct).\\
Now we list some properties of the matrices $A^{\nu}, \nu=
k_{1}<k_{2} \cdots <k_{n}$:
\begin{equation}
(a) \ \ \ A^{\nu}(\pi, \pi ) = 1,
\end{equation}
\begin{equation}
(b) \ \ \ A^{\nu}(\sigma, \pi ) = A^{\nu}(\pi, \sigma )^{*}\ ({\rm
hermiticity})
\end{equation}
(c) \ Let $w_{n} = n\dots 2 1$ be the longest permutation in
$S_{n}$. Then
\begin{equation}
A^{\nu}(\pi w_{n},\sigma w_{n}) = A^{\nu}(\sigma,\pi) =
A^{\nu}(\pi, \sigma)^{*}
\end{equation}
The property c) can be rewritten in the matrix form as follows :
\begin{equation}
W A^{\nu} W = (A^{\nu})^{T},\ \ W^{2}=1,
\end{equation}
where
\begin{equation}
W(\pi, \sigma ) = \left\{
\begin{array}{l}
1, \mbox{ if } \pi w_{n}=\sigma\\
0, \mbox{ otherwise}
\end{array}
\right.
\end{equation}
It is important to note that the Fock space, in our case, is positive
definite iff the Gram matrix  $A$ is positive definite. Recall that  a sufficient condition
for the positivity of norm squared of all vectors is (cf [9, Theorem 1.9.4])
\begin{equation}
|q_{ij}| < 1, \ \ 
 {\forall} \ i,j \in I.
\end{equation}
In particular, the condition (12) implies that the $n$--particle
states $|{\bf i}\rangle=a_{i_{1}}
^{\dagger}\cdots a_{i_{n}}^{\dagger}|0\rangle$ $(i_{j}\in I, n\geq 0$) are linearly independent.\\
{\bf Examples}: For the generic weights ${\nu}$=1,12,123 the Gram matrices are as follows:
$$A^{1}=(1); \ \ A^{12}=\left( \begin{array}{@{}c@{}c@{}}
1&q_{12}\\
q_{21}&1
\end{array}\right);$$
$$ A^{123} = \begin{array}{|c|c|@{}c|@{}c|@{}c|@{}c|@{}c|@{}}\hline
\lover{3}{\pi}\! \diagdown \lover{-2}{\sigma}&123&132&312&321&231&213\\ \hline
123&1 & q_{23} & q_{13}q_{23} & q_{12}q_{13}q_{23} & q_{12}q_{13}
& q_{12}
\\ \hline
132&q_{32} & 1 & q_{13} & q_{12}q_{13} & q_{12}q_{13}q_{32} &
q_{12}q_{32} \\ \hline 312&q_{31}q_{32} & q_{31} & 1 & q_{12} &
q_{12}q_{32} & q_{12}q_{31}q_{32} \\ \hline 321&q_{21}q_{31}q_{32}
& q_{21}q_{31} & q_{21} & 1 & q_{32} & q_{31}q_{32} \\ \hline
231&q_{21}q_{31} & q_{21}q_{31}q_{23} & q_{21}q_{23} & q_{23} & 1
& q_{31} \\ \hline
213&q_{21} & q_{21}q_{23} & q_{21}q_{13}q_{23}
& q_{13}q_{23} & q_{13} & 1 \\\hline
\end{array}$$ \\
(here we use the Johnson-Trotter ordering of permutations:123,132,312,321,231,213).
\\
For the non-generic: $\tilde{\nu}$=11, 113, the Gram matrices are:
$$A^{11}=(1+q_{11}); \ \  \ A^{113} = \begin{array}{|c|c|c|c|}\hline
\lover{3}{\pi}\! \diagdown \lover{-2}{\sigma}&113&131&311\\ \hline 113&1+q_{11} &
q_{13}+q_{11}q_{13} & q_{13}^{2}+q_{11}q_{13}^{2} \\ \hline
131&q_{31}+q_{11}q_{31} & 1+q_{11}q_{13}q_{31} &
q_{13}+q_{11}q_{13} \\ \hline
311&q_{31}^{2}+q_{11}q_{31}^2 &
q_{31}+q_{11}q_{31} & 1+q_{11}
\\\hline
\end{array} $$ \\
The inverses of the Gram matrices in the generic case above, are
given by:
$$[A^{12}]^{-1}=\frac{1}{\Delta^{12}}\left(\begin{array}{@{}c@{}c@{}}
1&-q_{12}\\-q_{21}&1
\end{array}\right)=\frac{1}{\Delta^{12}}\left(\begin{array}{@{}c@{}c@{}}
1&q_{12}\\
q_{21}&1
\end{array}\right)*\left(\begin{array}{@{}c@{}c@{}}
1&-1\\
-1&1
\end{array}\right)$$
where
$ \Delta^{12}:=1-q_{12}q_{21}=1-|q_{12}|^2 $, and
$$ [A^{123}]^{-1}=\frac{1}{\Delta^{123}}\ A^{123}\ast M^{123}$$
Here
$\Delta^{123}:=(1-|q_{12}|^2)(1-|q_{13}|^2)(1-|q_{23}|^2)(1-|q_{12}|^2|q_{13}|^2|q_{23}|^2)$,
$\ast$ denotes the Schur product  of matrices $
(a_{ij})\ast (b_{ij}) :=(a_{ij}b_{ij})$ \ and $M^{123}$ stands for
the following matrix:
{\footnotesize
$$\begin{array}{|c|@{}|@{}c|@{}c|@{}c|@{}c|@{}c|@{}c|@{}}\hline
\lover{3}{\pi}\! \diagdown \lover{-2}{\sigma}&123&132&312&321&231&213\\ \hline 123&\
(1\!-\!ac)(1\!-\!b)\!&(b\!-\!1)(1\!-\!c)\!&\
c(b\!-\!1)(1\!-\!a)&(1\!-\!ac)(1\!-\!b)&\ a(b\!-\!1)(1\!-\!c)&\
(b\!-\!1)(1\!-\!a)\\\hline 132&\ (c\!-\!1)(1\!-\!b)\!&\
(1\!-\!ab)(1\!-\!c)\!&\ (c\!-\!1)(1\!-\!a)&\ a(c\!-\!1)(1\!-\!b)&\
(1\!-\!ab)(1\!-\!c)&\ b(c\!-\!1)(1\!-\!a)\\\hline
312&\ (a\!-\!1)(1\!-\!b)\!&\ (a\!-\!1)(1\!-\!c)\!&\
(1\!-\!bc)(1\!-\!a)&\ (a\!-\!1)(1\!-\!b)&\ b(a\!-\!1)(1\!-\!c)&\
(1\!-\!bc)(1\!-\!a)\\\hline 321&\ (1\!-\!ac)(1\!-\!b)\!&\
a(b\!-\!1)(1\!-\!c)\!&\ (b\!-\!1)(1\!-\!a)&\ (1\!-\!ac)(1\!-\!b)&\
(b\!-\!1)(1\!-\!c)&\ c(b\!-\!1)(1\!-\!a)\\\hline 231&\
a(c\!-\!1)(1\!-\!b)\!&\ (1\!-\!ab)(1\!-\!c)\!&\
b(c\!-\!1)(1\!-\!a)&\ (c\!-\!1)(1\!-\!b)&\ (1\!-\!ab)(1\!-\!c)&\
(c\!-\!1)(1\!-\!a)\\\hline 213&\ (a\!-\!1)(1\!-\!b)\!&\
b(a\!-\!1)(1\!-\!c)\!&\ (1\!-\!bc)(1\!-\!a)&\
c(a\!-\!1)(1\!-\!b)&(a\!-\!1)(1\!-\!c)&\
(1\!-\!bc)(1\!-\!a)\\\hline
\end{array}$$ }
(with  $ a:=|q_{23}|^2, b:=|q_{13}|^2, c:=|q_{12}|^2$).\\
The inverse in the non-generic case $\nu =113$ is given by
$$ [A^{113}]^{-1}=\frac{1}{\Delta^{113}}\left(\begin{array}{@{}ccc@{}}
1 & -(1+q_{11})q_{13}&q_{11}q_{13}^2\\
-q_{31}(1+q_{11}) & (1+q_{11})(1+q_{13}q_{31}) & -(1+q_{11})q_{13}\\
q_{31}^2q_{11} & -q_{31}(1+q_{11}) & 1
\end{array}\right)$$
where
$\Delta^{113}=(1+q_{11})(1-q_{13}q_{31})(1-q_{11}q_{13}q_{31})
=(1+q_{11})(1-|q_{13}|^2)(1-q_{11}|q_{13}|^2).$

\noindent {\bf 3. Series expansions of number operators.}\\
First we recall that the $k$-th particle number operator  $N_{k} (k \in
I)$  (in the Fock representation satisfying the positivity
condition (12)) is a diagonal operator which  counts the number of appearances
of the creation operator $a_{k}^{\dagger}$ in any multi-particle state
 $|{\bf i}\rangle$.
These operators satisfy the following implicit conditions (equations):
\begin{equation}
\begin{array}{c}
[N_{k},a_{l}]= -a_{k}\delta_{kl}, \ \ \forall k,l \in I \\
N_{k}|0\rangle = 0,\ \ \forall k \in I
\end{array}
\end{equation}

Note that for any fixed $k\in I$, if we assume (12), the equations
(13) have unique  solution for $N_{k}$.\
The number operators play an important role in constructing the
free Hamiltonian (= the energy operator) of the free system (for
which the energy is additive, cf.\ Greenberg [3]) of generalized
quon particles in the nonrelativistic limit:
\begin{equation}
H = \sum_{k\in I}E_{k}N_{k}
\end{equation}
More generally, our primary goal here is to express $N_{k}$ in
terms of quon algebra generators
as an normally ordered infinite  series involving certain iterated
deformed commutators of the creation and annihilation operators.

It is already indicated in [8] that the formal expansion of the
number operator $N_{k}$ in terms of normally ordered products is
necessarily of the following form which preserves each
$n$-particle subspace (it easily follows from (3)):
\begin{equation}
N_{k} = \sum_{{\bf i}\in I^{+},\ i_{1}=k}X_{{\bf i}}^{\dagger}Y_{{\bf i}}
\end{equation}
where $I^{+}$ denotes the set of all nonempty words (or sequences)
${\bf i}=i_{1}\dots i_{n}$, $n\geq 1$ over the set $I$ as an
alphabet, and the sum is over those  words which begin with letter
$k$. Here, if the indices  $i_{1}, \dots, i_{n}$ are distinct, we
require that  $X_{{\bf i}}$ and $Y_{{\bf i}}$ are both
multihomogeneous of the same multidegree, i.e.\ they are
expressible as a  linear combination of all  rearrangements
$a_{\bf j}=a_{\bf i}.{\pi} :=a_{{\bf
i}.{\pi}}(=a_{i_{{\pi}(1)}}a_{i_{{\pi}(2)}}\cdots
a_{i_{{\pi}(n)}}) $ of the ``monomial'' $a_{\bf i} =
a_{i_{1}}a_{i_{2}}\cdots a_{i_{n}}$, in the following form:
\begin{equation}
X_{{\bf i}} = \sum_{\pi \in S_{n}}a_{{\bf i}.{\pi}}x_{{\bf i}.{\pi},{\bf i} }
\end{equation}
\begin{equation}
Y_{{\bf i}} = \sum_{\pi \in S_{n}}a_{{\bf i}.{\pi}}y_{{\bf
i}.{\pi}, {\bf i}}
\end{equation}
\noindent where $x_{{\bf i}.\pi,{\bf i}}$ and $y_{{\bf i}.\pi,{\bf
i}}$ are, yet unknown, coefficients (depending on $q_{ij}$'s) with
the following normalization convention  $y_{{\bf i},  {\bf i}}
=1$. For general {\bf i}' s, we require that the summations in
(16) and (17) should be replaced by summations over the left
cosets $H\backslash S_{n}$,  where $H={\rm Stab}_{{\bf i}}$ is the
stabilizer subgroup of $S_{n}$ fixing ${\bf i}$, with coefficients
$\tilde{x}_{{\bf i}.\tilde{\pi},{\bf i}}$, $\tilde{y}_{{\bf
i}.\tilde{\pi}, {\bf i}}$,
$\tilde{\pi}\in  H\backslash S_{n}$ equal to the following orbit sums:
\begin{equation}
\begin{array}{c}
\displaystyle \tilde{x}_{{{\bf i}.\tilde{\pi}},{\bf i}} = \sum_{h\in H}x_{{{\bf i}.h\pi},{\bf i}}\\
\displaystyle \tilde{y}_{{{\bf i}.\tilde{\pi}},{\bf i}} = \sum_{h\in H}y_{{{\bf i}.h\pi},{\bf i}}
\end{array}
\end{equation} \\
\indent Now we start finding the solution of the system (13), in
the form (15), as follows: We first use the fact that under the
condition (12), the set of all monomials
$a_{i_{n}}^{\dagger}\cdots a_{i_{1}}^{\dagger} a_{j_{1}}\cdots
a_{j_{m}}$, ($i_{k}, j_{l}\in I$) is linearly independent.Then, we
plug the right hand side of (15) into the system(13). By resolving
it successively  in degree one, then in degree two, etc.,
we obtain the following (noncommutative) recursions for~$Y_{{\bf i}}$'s:\\
{\bf RECURSIONS \  FOR \  $Y$'s}:\vspace{-5mm}
\begin{equation}
Y_{i_{1}i_{2}\cdots i_{n}} = Y_{i_{1}\cdots i_{n-1}}a_{i_{n}} -
q_{i_{n}i_{1}}q_{i_{n}i_{2}}\cdots
q_{i_{n}i_{n-1}}a_{i_{n}}Y_{i_{1} \cdots i_{n-1}}
\end{equation}
\noindent and similarly, a system of ``twisted'' partial
differential equations for  $X_{{\bf i}}$'s:\\
{\bf EQUATIONS \ FOR \ $X$'s}:\vspace{-5mm}
\begin{equation}
{}_{l}\partial (X_{i_{1}\cdots i_{n}})^{\dagger} = (X_{i_{1}\cdots
i_{n-1}})^{\dagger}\delta_{li_{n}}\ \ \ (l\in \{i_{1},\dots
,i_{n}\})
\end{equation}
\noindent
where ${}_{l}\partial $ denotes the left twisted derivative:
\begin{equation}
{}_{l}\partial (a_{j_{1}}^{\dagger}\cdots a_{j_{n}}^{\dagger}) =
\sum_{(p:j_{p}=l)}q_{lj_{1}}\cdots q_{lj_{p - 1}}a_{j_{1}}^{\dagger}
\cdots \widehat{{a}_{j_{p}}^{\dagger}}\cdots
a_{j_{n}}^{\dagger}
\end{equation}
(\  $\widehat{} $\ \  denotes the omission of the corresponding creation operator).
\bprop
The Y-components (17) of the solution (15) of  eq.(13) are given by the following iterated
{\bf q}-commutator (``generalized Lie elements'') formula:
\begin{equation}
\begin{array}{c}
Y_{i_{1}}=a_{i_{1}};\\
Y_{i_{1}i_{2}\dots i_{n}} = [\cdots
[[a_{i_{1}},a_{i_{2}}]_{q_{i_{2}i_{1}}},
a_{i_{3}}]_{q_{i_{3}i_{1}}q_{i_{3}i_{2}}},\dots,
a_{i_{n}}]_{q_{i_{n}i_{1}} q_{i_{n}i_{2}}\cdots q_{i_{n}i_{n-1}}}
\end{array}
\end{equation}
\noindent where $[x,y]_{q}=xy - qyx$ denotes the $q$-commutator of
$x$ and $y$.(For $N_{k}$ we need to set $i_{1}=k$).
 \eprop
{\it Proof}: By iterating (19).\\
In order to express the formula (22)(and some others later) in the
operator  form we shall now introduce a twisted group algebra of
the permutation group.\ \

 \noindent {\bf 4.\ A twisted group algebra action.} \\ Let us
 consider  \\
 (1)A {\it right action} of the symmetric group $S_{n}$, by permuting
factors of any degree n monomial in the annihilation operators:
\begin{equation}
a_{\bf i}.{ \pi}=(a_{i_{1}}a_{i_{2}}\cdots a_{i_{n}}). \pi :=
a_{i_{\pi (1)}}a_{i_{\pi (2)}}\cdots a_{i_{\pi (n)}}
\end{equation}
(2)A {\it "diagonal" action} of the formal power series ring $K_{n}={\bf C}
[[Q_{k,l}, 1\leq k
, l\leq n]]$\\
(where $Q_{k,l}$ are commuting indeterminates) defined by:
\begin{equation}
a_{{\bf i}}.Q_{k,l}(=(a_{i_{1}}a_{i_{2}}\cdots a_{i_{n}}).Q_{k,l}):=
q_{i_{k}i_{l}}a_{i_{1}}a_{i_{2}}\cdots a_{i_{n}}
\end{equation}
(here $q_{ij}$\ 's are complex numbers from the canonical commutation relations (1)!).
These two actions give rise to an action of a {\it twisted group algebra\/}:
\begin{equation}
{\cal K}_{n} =K_{n}\tilde{ \ }[S_{n}]
\end{equation}
of $S_{n}$ (with coefficients in $K_{n}$).The multiplication in
the algebra $ {\cal K}_{n}$ is defined by imposing the following
comutation relations (``an action of $S_{n}$ on the coefficient
ring $K_{n}$ '')
\begin{equation}
\pi Q_{k,l}=Q_{\pi (k)\pi (l)}\pi
\end{equation}
It is clear that, by specializing $Q_{k,l}=q$ \ ( $1\leq k, l\leq
n$), the twisted group algebra $K_{n}\tilde{ \ }[S_{n}]$ is mapped
onto the ordinary group algebra ${\bf C}[[q]][S_{n}]$ in which,
according to Zagier [15], live certain important elements :
 $\alpha_{n}$, $\beta_{n}$, $\gamma_{n}$, $\delta_{n}$ satisfying
\begin{equation}
\alpha_{n}=\alpha_{n-1}\beta_{n}, \beta_{n}=\delta_{n}{\gamma_{n}}^{-1}
(\Rightarrow
\alpha_{n}=\beta_{2}\cdots \beta_{n}=\delta_{2}\gamma_{2}^{-1}\delta_{3}
\gamma_{3}^{-1}\cdots \gamma_{n-1}^{-1}\delta_{n}\gamma_{n}^{-1})
\end{equation}
(Note that our notation for $\delta_{n}$ is shifted by 1 compared with [15], which seems to be more natural!)\\
These elements, via the regular representation $R_{n}$, were crucial in Zagier's computation of the determinant and the inverse of the one-parameter matrices $A_{n}=A_{n}(q)=R_{n}(\alpha_{n}).$
We shall now define a  ``lifting'' to $K_{n}\tilde{\ }[S_{n}]$ of  the Zagier's elements
by first defining, for each permutation $\pi \in S_{n}$, an element
$\tilde{\pi}\in K_{n}\tilde{\ }[S_{n}]$,($\pi \in S_{n}$), which encodes all inversions of $\pi$:
\begin{equation}
\begin{array}{l}
\tilde{\pi}:  = Q_{\pi}\pi, \quad
\mbox{where}  \quad  Q_{\pi}:=\prod_{1\leq k<l\leq n \
,\pi(k)>\pi(l)}Q_{\pi(k),\pi(l)}\\[3mm]
\hspace{-2.5cm}\mbox{with the multiplication rule}\\[3mm]
 \tilde{\sigma}\tilde{\pi}=(\prod_{(a,b)\in
I(\sigma )\cap
I(\pi^{-1})}Q_{\sigma(a),\sigma(b)}Q_{\sigma(b),\sigma(a)})
\widetilde{\sigma \pi} \, .
\end{array}
\end{equation}
\noindent
(Observe that $\tilde{\pi}$ generalizes $q^{i(\pi)}\pi $, $i(\pi):=$the number of inversions
of $\pi $).\\
Then we define a ``lifting'' of all Zagier's elements by the following formulas:
\begin{eqnarray}
\tilde{\alpha}_{n}:&=&\sum_{\pi \in S_{n}}\tilde{\pi},\\
\tilde{\beta}_{n}:&=&\sum_{k=1}^{n}\tilde{t}_{k,n}\\
\tilde{\gamma_{n}}:&=&(1-\tilde{t}_{1,n})(1-\tilde{t}_{2,n})\cdots (1-\tilde{t}_{n-1,n})\\
\tilde{\delta}_{n}:&=&(1-\tilde{t}_{n-1}\tilde{t}_{1,n})(1-\tilde{t}_{n-1}\tilde{t}_{2,n})\cdots
(1-\tilde{t}_{n-1}\tilde{t}_{n-1,n})
\end{eqnarray}
Similarly we define\\
\null\hfill$\displaystyle
\tilde{\alpha}_{n_1,n_2,\ldots,n_k}:=\sum_{\pi\in S_{n_1}\times
S_{n_2}\times\cdots\times S_{n_k}}\tilde{\pi}$\hspace{1cm}\hfill (29a)

\noindent (Here $t_{k,l}$ denotes the cycle $\left(
                   \begin{array}{cccc}
                   k & k+1 & \cdots & l \\
                   l & k   & \cdots & l-1
                   \end{array}
                   \right)\in S_{n}$ and $t_{k}:=t_{k,k+1}$.)\\
It is easy to check that the following relations,  analogous to (27),
hold true:
\begin{equation}
\tilde{\alpha}_{n}=\tilde{\alpha}_{n-1}\tilde{\beta}_{n},\tilde{ \beta}_{n}=\tilde{\delta}_{n}\tilde{\gamma}_{n}^{-1}
(\Rightarrow \tilde{\alpha}_{n}=\tilde{\beta}_{2}\cdots \tilde{\beta}_{n}=
\tilde{\delta}_{2}{\tilde{\gamma}_{2}}^{-1}\tilde{\delta}_{3}
{\tilde{\gamma_{3}}}^{-1}\cdots {\tilde{\gamma}_{n-1}}^{-1}
\tilde{\delta}_{n}
\tilde{\gamma_{n}}^{-1})
\end{equation}

{\bf {Important note}}. Now we can realize all Gram matrices $A^{\nu}$ from (4) as the matrices of the right multiplication by the
lifted Zagier element $\tilde{\alpha}_{n}$ on the space monomials $a_{{\bf i}}$ of weight $\nu$.This explains why we needed
to introduce a twisted group algebra in the multiparameter case.

In what follows, we shall also need the following notations:
\begin{eqnarray}
Q_{\{ \pi \} }:&=&\prod_{1\leq k<l\leq n,\ \pi(k)>\pi(l)}Q_{\pi(k),\pi(l)}Q_{\pi(l),\pi(k)} \ \ ({\rm for \  any}\ \ \pi\in S_{n}),\\
Q_{T}:&=&\prod_{k\neq l\in T}Q_{k,l} \ \ ({\rm  for \ any \ set }\
T\subseteq \{ 1,2,\dots, n\}  )
\end{eqnarray}\\
together with the following Lemma which we shall use in the proof of the main result:
\blemma
We have the following identity in ${\cal K}_{n}$:
\begin{equation}
\tilde{\alpha}_{n-1,1}(1-\tilde{t}_{n-1}\tilde{t}_{1,n})=\xi_n\tilde{\alpha}_{1,n-2,1},
\end{equation}
where $\xi_n:=\sum_{k=1}^{n-1}(1-Q_{\{ k,k+1\} }\cdots Q_{\{ k,n\} })
\tilde{t}_{1,k}$.\\
(Recall from (29a) that
$\tilde{\alpha}_{n-1,1}=\sum_{\pi \in S_{n-1}\times
S_{1}}\tilde{\pi}$, $\tilde{\alpha}_{1,n-2,1}=\sum_{\pi \in
S_{1}\times S_{n-2}\times S_{1}}\tilde{\pi}$.)
\elemma
{\it Proof\/}:
By definition  $\tilde{\alpha}_{n-1,1}=\sum_{\pi \in S_{n-1}\times
S_{1}}\tilde{\pi}$. By using a factorization $\pi =t_{1,k}\sigma
$, where $\pi(1)=k$, $\sigma \in S_{1}\times S_{n-2}\times S_{1}$,
we get $\tilde{\alpha}_{n-1,1}=
(\sum_{k=1}^{n-1}\tilde{t}_{1,k})\tilde{\alpha}_{1,n-2,1}$ (here
we used that $\tilde{\pi}=\tilde{t}_{1,k}\tilde{\sigma}$,
c.f.(28)). Similarly,
\begin{eqnarray*}
\tilde{\alpha}_{n-1,1}\tilde{t}_{n-1}\tilde{t}_{1,n} & = &
\sum_{\pi \in S_{n-1}\times S_{1}}\tilde{\pi}\tilde{t}_{n-1}\tilde{t}_{1,n}=
\sum_{\pi \in S_{n-1}\times S_{1}}\tilde{\pi}Q_{\{ n-1,n\} }\tilde{t}_{1,n-1} \ \ (by (28)) \\
          & = &
\sum_{\pi \in S_{n-1}\times S_{1}}Q_{\{ \pi(n-1),\pi(n)\} }\tilde{\pi}
\tilde{t}_{1,n-1} \\
          & = &
\sum_{\pi \in S_{n-1}\times S_{1}}Q_{\{ \pi(n-1),n\} }Q_{\{ t_{\pi(n-1),n-1}^{-1}\} }
\widetilde{\pi t}_{1,n-1} \ \ (by (28) and (34)) \\
          & = &
\sum_{\sigma\in S_{1}\times S_{n-2}\times S_{1}}Q_{\{ t_{\pi(n-1),n}^{-1}\} }\tilde
{t}_{1,\pi(n-1)}\tilde{\sigma} \,[t_{1,\pi(n-1)} \sigma=\pi t_{1,n-1}]\\
          & = &
\left(\sum_{k=1}^{n-1}Q_{\{{ t_{k,n}^{-1}\}}
}\tilde{t}_{1,k}\right)\tilde{\alpha}_{1,n-2,1}.
\end{eqnarray*}
By subtracting the last two formulas,
the Lemma follows.\\
Now we state the formula (22) in the operator form:
\bcor
We have\\
\noindent i) $ Y_{i_{1}\cdots i_{n}}=(a_{i_{1}}a_{i_{2}}\cdots
a_{i_{n}}). \overline{ \gamma_{n}}$, where \  \
$\overline{\gamma_{n}}:=(1-\widetilde{t}_{1,2})(1-\widetilde{t}_{1,3})\cdots (1-\widetilde{t}_{1,n})\in {\cal K}_{n}. $ \\
ii) $a_{i_{1}}a_{i_{2}}\cdots a_{i_{n}}=Y_{i_{1}\cdots i_{n}}.\overline{ \gamma
_{n}}^{-1}$, with
$$\overline{\gamma_{n}}^{-1}=\sum_{\pi\in S_{n}}\tilde{\pi}\cdot \prod_{\pi (i)>\pi (i+1)}
Q_{\{ 1,\dots, i\} }/(1-Q_{\{ 1,2\} })\cdots (1-Q_{\{ 1,\dots, n\}
})$$ iii) The set $\{ Y_{{\bf i}\cdot \tilde{\pi}}| \tilde{\pi}\in
H\backslash S_{n}\} $, $(H=St_{{\bf i}})$ is a linearly
independent set if $|q_{i_{r}i_{s}}|
<1$, $1\leq r\neq s\leq n$.
\ecor
{\it Proof}. i) The formula (22) can be rewritten as
$$Y_{i_{1}i_{2}\cdots i_{n}}=a_{i_{1}}a_{i_{2}}\cdots a_{ i_{n}}(1-q_{i_{2}i_{1}}t_{1,2})(1-q_{i_{3}i_{1}}q_{i_{3}i_{2}}t_{1,3})\cdots (1-q_{i_{n}i_{1}}q_{i_{n}i_{2}}
\cdots q_{i_{n}i_{n-1}}t_{1,n})$$
By using  $\widetilde{t}_{1,l}=Q_{l,1}\cdots Q_{l,l-1}t_{1,l}=t_{1,l} Q_{1,2}Q_{1,3}\cdots Q_{1,l}$\, the claim i) follows.\\
ii) The proof of ii) is similar to that of Proposition.2.1.1 in [10].\\
iii) Follows from ii).
\bprop
The $Y_{{\bf i}}$'s satisfy the following (twisted) differential equations:
\begin{equation}
\begin{array}{rr@{\ =\ }l}
i)&  {}_{l}\partial (Y_{i_{1}\cdots i_{n}})^{\dagger}&\sum_{(j\geq 2:i_{j}
=l)}d^{(j)}_{i_{1}\cdots i_{n}}(Y_{i_{1}\cdots \hat{i_{j}}\cdots i_{n}})
^{\dagger},\ \ ( n\geq 2) \\
\\
ii)&  {}_{l}\partial Y^{\dagger}_{i_{1}}&\delta_{i_{1}l},\ \ ( n=1)
\end{array}
\end{equation}
where ${}_{l}\partial $ is defined in (21), and where
\begin{equation}
d^{(j)}_{i_{1}\cdots i_{n}}:=q_{i_{j}i_{j+1}}\cdots q_{i_{j}i_{n}}
(1-|q_{i_{j}i_{1}}\cdots q_{i_{j}i_{j-1}}|^{2})
\end{equation}
\eprop
{\it Proof}. By induction. For $n=2$ we have $Y_{i_{1}i_{2}}=[a_{i_{1}},a_{i_{2}}]
_{q_{i_{2}i_{1}}}=a_{i_{1}}a_{i_{2}}-q_{i_{2}i_{1}}a_{i_{2}}a_{i_{1}}$
 what implies $(Y_{i_{1}i_{2}})^{\dagger}=a_{i_{2}}^{\dagger}a_{i_{1}}^{\dagger}-q_{i_{1}i_{2}}
a_{i_{1}}^{\dagger}a_{i_{2}}^{\dagger}$ (here we use
$({q}_{ij})^*=q_{ji}$). Hence
$${}_{l}\partial (Y_{i_{1}i_{2}})^{\dagger}=\delta_{li_{2}}a_{i_{1}}^{\dagger}
+\delta_{li_{1}}q_{li_{2}}a_{i_{2}}^{\dagger}-q_{i_{1}i_{2}}(\delta_{li_{1}}
a_{i_{2}}^{\dagger}+\delta_{li_{2}}q_{li_{1}}a_{i_{1}}^{\dagger})$$
$$=(1-q_{i_{1}i_{2}}q_{i_{2}i_{1}})a_{i_{1}}^{\dagger}\delta_{li_{2}}=
d^{(2)}_{i_{1}i_{2}}Y^{\dagger}_{i_{1}}\delta_{li_{2}}$$
Now we suppose that (25 ) holds true for $n-1$. Then, from (19) it follows that

$${}_{l}\partial (Y_{i_{1}\cdots i_{n}})^{\dagger}={}_{l}\partial
[a_{i_{n}}^{\dagger}(Y_{i_{1}\cdots i_{n-1}})^{\dagger}-q_{i_{1}i_{n}}
\cdots q_{i_{n-1}i_{n}}(Y_{i_{1}\cdots i_{n-1}})^{\dagger}a_{i_{n}}^{\dagger}]
$$
$$=\delta_{li_{n}}(Y_{i_{1}\cdots i_{n-1}})^{\dagger}+q_{li_{n}}a_{i_{n}}^
{\dagger}{}_{l}\partial (Y_{i_{1}\cdots i_{n-1}})^{\dagger}$$
$$-q_{i_{1}i_{n}}\cdots q_{i_{n-1}i_{n}}[{}_{l}\partial (Y_{i_{1}\cdots
i_{n-1}})^{\dagger}a_{i_{n}}^{\dagger}+q_{i_{n}i_{1}}\cdots q_{i_{n}i_{n-1}}
\delta_{li_{n}}(Y_{i_{1}\cdots i_{n-1}})^{\dagger}]$$
$$=\delta_{li_{n}}(1-|q_{i_{1}i_{2}}\cdots q_{i_{n-1}i_{n}}|^{2})(Y_{i_{1}
\cdots i_{n-1}})^{\dagger}+ $$
$$\sum_{j=2;i_{j}=l}^{n-1}q_{li_{n}}d^{(j)}_{i_{1}\cdots i_{n-1}}
[a_{i_{n}}^{\dagger}(Y_{i_{1}\cdots \widehat{i_{j}}\cdots i_{n-1}})^{\dagger}
-q_{i_{1}i_{n}}\cdots \widehat{{q}_{li_{n}}}\cdots q_{i_{n-1}i_{n}}
(Y_{i_{1}\cdots \widehat{i}_{j}\cdots i_{n-1}})^{\dagger}a_{i_{n}}^{\dagger}]$$
$$=\delta_{li_{n}}d^{(n)}_{i_{1}\cdots i_{n}}(Y_{i_{1}\cdots i_{n-1}})^{\dagger}+
\sum_{j=2;i_{j}=l}^{n-1}d^{(j)}_{i_{1}\cdots i_{n}}(Y_{i_{1}\cdots \widehat{i}
_{j}\cdots i_{n}})^{\dagger} $$
$$=\sum_{n\geq j\geq 2; i_{j}=l}d^{(j)}_{i_{1}\cdots i_{n}}
(Y_{i_{1}\cdots \widehat{i}_{j}\cdots i_{n}})^{\dagger}.$$
This completes the proof of Proposition 2.
\\
\indent
Now we proceed with solving (20) to get $X_{{\bf i}}$-components of
our number operator $N_{k}$. There are two approaches: \\
\indent The first approach, developed in [8], is based on an
observation that in (37) the index $i_{1}$ survives in all terms
of the r.h.s. So, we could look for $X_{{\bf i}}$'s in the form of
a linear combination of such $Y_{{\bf i}}$'s with the first index
fixed ($=k$ for $N_{k}$).
\begin{equation}
(X_{{\bf i}})^{\dagger}=\sum_{
                  {\bf j}={\bf i}.\pi, \pi \in S_{1}\times S_{n-1}}
                   (Y_{{\bf j}})^{\dagger}c_{{\bf j},{\bf i}}
\end{equation}
By applying the twisted derivative ${}_{l}\partial $ to (39), the left hand
side gives
$${}_{l}\partial (X_{{\bf i}})^{\dagger}=(X_{i_{1}\cdots i_{n-1}})^{\dagger}
\delta_{li_{n}}\ \ (by (20))$$
$$=\sum_{\sigma \in S_{1}\times S_{n-2}}(Y_{i_{\sigma(1)}\cdots i_{\sigma(n-1)}})^{\dagger}
c_{i_{\sigma(1)}\cdots i_{\sigma(n-1)},i_{1}
\cdots i_{n-1}}
\delta_{li_{n}}\ \ (by (39)).$$
The ${}_{l}\partial $ applied to the right hand side of (39) gives:
$$\sum_{\pi \in S_{1}\times S_{n-1}}
{}_{l}\partial (Y_{{\bf i}. \pi})^{\dagger}c_{{\bf i}\cdot
\pi,{\bf i} }$$
$$=\sum_{\pi \in S_{1}\times S_{n-1}}
\sum_{(n\geq j\geq 2; l=\pi(j))}d^{(j)}_{{\bf i}. \pi}
(Y_{i_{\pi(1)}\cdots \widehat{i_{\pi(j)}}\cdots
i_{\pi(n)}})^{\dagger}c_{{\bf i}. \pi,{\bf i}}\ \ (by (37)).$$ By
linear independence of $Y_{{\bf i}}$'s (cf. Corollary 1) we obtain
the following system of $(n-1)!$
 equations (in the generic case) for $(n-1)!$ unknown coefficients
$c_{{\bf i}. \pi,{\bf i}}$, ($i_{1}=k, \pi \in S_{1}\times S_{n-1}$):\\
{\bf EQUATIONS \ FOR} \ ${c_{{\bf j},{\bf i}}}$'s:\vspace{-5mm}
\begin{equation}
\sum_{n\geq j\geq 2}d^{(j)}_{{\bf i}.\pi t_{j,n}} c_{{\bf i}.\pi
t_{j,n},{\bf i}} =\delta_{\pi(n),n}c_{({\bf i}.\pi)^{'},{\bf
i}^{'}}
\end{equation}
where $\pi \in S_{1}\times S_{n-1}$, $t_{j,n}$ denotes the cyclic
permutation which sends $1,2,\dots, j,j+1,\dots, n$ to $1,2,\dots,
n,j, \dots, n-1$ and ${\bf i}^{'}=i_{1}\dots i_{n-1}$.\\
Note that our derivation of the equations (40) (generic case) will
yield (by summation) the equations for the nongeneric case (i.e.
when there are repetitions among $ i_{1},\cdots, i_{n}\ 's)$. This
justifies the form of our
 expression (15) for the number operators $ N_{k}^{\ ,}s$.\\
\indent The second approach to solving the recursive system (20)
for $X_{{\bf i}}$'s is to write $Y_{{\bf i}}$'s in terms of
$X_{{\bf i}}$'s, again with the first index fixed ($=k$ for
$N_{k}$).
\begin{equation}
(Y_{{\bf i}})^{\dagger}=\sum_{{\bf j}={\bf i}.\pi, \pi \in
S_{1}\times S_{n-1}} (X_{{\bf j}})^{\dagger}e_{{\bf j},{\bf i}}
\end{equation}
\bprop
The coefficients $e_{{\bf j},{\bf i}}$ satisfy the
following recursions:
\begin{equation}
e_{{\bf i}.\pi,{\bf i}}=d_{{\bf i}}^{(r)}e_{{\bf i}^{'}.\pi^{'},{\bf i}^{'}}
\end{equation}
where\\
$r=\pi(n)$, ${\bf i}^{'}=i_{1}\dots i_{n-1}$, $\pi^{'}= t_{r,n}\pi (\Rightarrow \pi=t_{r,n}^{-1}\pi^{'},\pi^{'} \in S_{n-1}) $,\\
and $d_{{\bf i}}^{(r)}=d_{i_{1}\cdots i_{n}}^{(r)}$ is defined in (38).
\eprop
{\it Proof}. By applying ${}_{l}\partial $ to both sides of (41), and using (37),
we obtain
\begin{eqnarray}
\sum_{{\bf j}={\bf i}.\pi,\pi \in S_{1}\times S_{n-1}}(X_{j_{1}\dots j_{n-1}})^{\dagger}e_{{\bf j}, {\bf i}}
\delta_{l,j_{n}}=\sum_{r\geq 2, i_{r}=l}d_{{\bf i}}^{(r)}
(Y_{i_{1}\dots \hat{i}_{r}\dots i_{n}})^{\dagger}\\
=\sum_{r\geq 2, i_{r}=l}d_{{\bf i}}^{(r)}\sum_{\sigma \in S_{1}\times
S_{n-2}}(X_{{\bf i}_{\hat{r}}.\sigma})
^{\dagger}e_{{\bf i}_{\hat{r}}.\sigma,{\bf i}_{\hat{r}}}
\end{eqnarray}
where ${\bf i}_{\hat{r}}:=i_{1}\dots i_{r-1}i_{r+1}\dots i_{n}$. Observe
that $i_{\pi(1)}\dots i_{\pi(n-1)}={\bf i}_{\hat{r}}.\sigma $ iff $r=
\pi(n)$ and $\sigma = t_{r,n}\pi$.\\
By equating the coefficients in (43) and (44) the proof of Proposition 3. follows.\\
Note that the recursion (42) corresponds to the  multiplication by the following element
(of the twisted group algebra):\\
{\null\hfill
{${\eta}_n:=\sum_{k=2}^{n}Q_{\{ k,k+1\}} \cdots Q_{\{ k,n\}}(1-Q_{\{ k,1\} }\cdots Q_{\{ k,k-1\} })
{\tilde{t}_{k,n}}^{-1}$}.
\hfill (42a)}

\indent Let $E = (e_{{\bf i},{\bf j}})$, with $i_{1}=j_{1}(=k)$
fixed,be the $(n-1)!\times (n-1)!$ transition matrix (in the
generic case), with entries $e_{{\bf i},{\bf j}}$ from (41). In
[8] the linear equations for the entries of $E^{-1}$ are
constructed for general $n$ and solved in special cases for
$n=1,2,3$. From these computations it was conjectured (in [8]) that $E^{-1}$
is related to the inverse of the Gram matrix $A$, see eq.\ (3);
here we prove this conjecture.

By comparing $\xi_n$ from (36) with $\eta_n$ from (42a) we get
\[
w_n\eta_nw_n=\xi_n
\]
and we deduce the following:
\blemma
The matrix $E$ is the matrix of the
right multiplication by the following element of our twisted group
algebra $K_{n}\tilde{\ }[S_{n}]$:
\begin{equation}
 w_{n}\widetilde{\alpha}_{n-1,1}
\widetilde{\delta}_{n}w_{n}.
\end{equation}
Here $w_{n}=n\ldots 21$ denotes the longest element in $S_{n}$.
\elemma\ \
 {\it Proof}. Follows by iteratively applying the result
of Lemma 1, using the definition (32) of $\tilde{\delta}_n$
 together with the recursions obtained in the
Proposition 3. \\
{\bf 5. The main results.} Now we prove the
following theorem:\ \
\bthm
The number operators in the multiparameter quon algebra ${\cal
A}^{({\bf q})}$  eq.~(1) are given, in the expanded form, by:

\begin{equation}
N_{k}=a_{k}^{\dagger}a_{k}+\sum_{n=1}^{\infty}\ \sum_{{\bf i},
i_{1}=k} \sum_{\pi \in S_{1}\times S_{n-1}}\hat{A}_{{\bf i},{\bf
i}.\pi}^{-1} (Y_{ {\bf i}.\pi})^{\dagger}Y_{{\bf i}}
\end{equation}
where the matrix $\hat{A}$ denotes the matrix obtained from  the Gram matrix
$A =$\\ $ \oplus_{n\geq 0} \oplus_{k_{1}\leq \cdots \leq k_{n}}A^{k_{1}\dots k_{n}}$ (described in (4)) by
replacing each block $A^{k_{1}\dots k_{n}} (k_{1}\leq \cdots \leq k_{n})$ with a specialized $n!\times n!$
block $A^{12\cdots n}|_{1\mapsto k_{1},2\mapsto k_{2}\cdots n\mapsto k_{n}}$
 and  $Y_{{\bf i}}$ are given by (22).

Or, in the reduced form, by:
\begin{equation}
N_{k}=a_{k}^{\dagger}a_{k}+\sum_{n=1}^{\infty}\sum_{{\bf
i},i_{1}=k} \sum_{\tilde{\pi}\in Stab_{\bf i}\backslash
S_{1}\times S_{n-1}} \tilde{A}_{{\bf i},{\bf i}.\tilde{\pi}}^{-1}(Y_{ {\bf
i}.\tilde{\pi}}) ^{\dagger}Y_{{\bf i}},
\end{equation}
where the reduction procedure is given with respect to the groups $S_1\times S_{n-1}$
(instead of $S_n$)
analogously to
the reduction procedure described in the text preceding (6).

\ethm
The proof of this theorem relies  on  one more lemma.
\blemma
We have \\
The $S_{1,n-2,1}$ -component of ${\tilde{\alpha}_{n}}^{-1}$\\ $ =$
\noindent The  $S_{1,n-2,1}$ -component of \
$\widetilde{\delta}_{n}^{-1} \times
\widetilde{\alpha}_{n-1,1}^{-1}$ \elemma
{\it Proof of Lemma 3.}:
This is a generalization of a Zagier's result [15]. Here we sketch
the proof. By observing that
$\widetilde{\alpha}_{n-1}=\widetilde{\alpha}_{n-1,1}$ we can write
(c.f.(33))
$$\widetilde{\alpha}_{n} =\widetilde{\alpha}_{n-1,1}\widetilde{\delta}_{n}
\widetilde{\gamma}_{n}^{-1}$$
\begin{equation}
\widetilde{\alpha}_{n}^{-1}=\widetilde{\gamma}_{n}\widetilde{\delta}_{n}^{-1}
\widetilde{\alpha}_{n-1,1}^{-1}
\end{equation}
\noindent where, according to (31),
\begin{equation}
\tilde{\gamma}_{n}=(1-\widetilde{t}_{1,n})(1-\widetilde{t}_{2,n})\cdots
(1-\widetilde{t}_{n-1,n}) =\sum_{k=1}^{n}(-1)^{n-k}\sum_{\pi \in
S_{n,k}} \tilde{\pi}^{-1}
\end{equation}
\noindent
with  $S_{n,k}\subset S_{n}$ denoting the set of all permutations such that  $\pi(1)<\cdots <\pi(k)=n>\cdots
>\pi(n)$.Note that $\widetilde {\delta}_{n}$ involves only permutations belonging to $S_{n-1}\times S_{1}$
(c.f.\ (32); for an explicit formula for the inverse of
$\widetilde {\delta}_{n}$ see Proposition 2.1.1. in [10]). Now it
is clear that only the trivial term in $ \tilde{\gamma}_{n}$ can
contribute to the
 $S_{1,n-2,1}$ -component  of $ \tilde{\alpha}_{n}^{-1}$.
The Lemma 3 is proved. This establishes the connection between $E^{-1}$ and the inverse $A^{-1}$ of
the Gram matrices.\\
{\it Proof  of Theorem 1}.By using Lemmas 2 and 3, together with
the symmetry property (9) and hermiticity (8) of the multiparameter
Zagier matrices, we obtain
$$X^{\dagger}_{{\bf i}}=\sum_{\pi \in S_{1}\times S_{n-1}}Y^{\dagger}_{ {\bf i}.\pi}A^{-1}_{{\bf i},{\bf i}.\pi}$$
in expanded form, and  similarly
$$X^{\dagger}_{{\bf i}}=\sum_{\tilde{\pi} \in H\backslash S_{1}\times S_{n-1}}Y^{\dagger}_{ {\bf i}.\tilde{\pi}}A^{-1}_{{\bf i},{\bf i}.\tilde{\pi}}$$
in reduced form.
 This completes the proof of Theorem 1.
The method for calculating the inverse of the matrix $A$ is explained
in  [9,Theorem 2.2.17].

\bcor
Let us assume infinite set $I$ of indices, then the number operator $N_k$
restricted to the finite subset $I_f\subseteq I$ is obtained from
eq.\ (46), eq.\ (47) by projecting out all words with letters
from the subset $I_f$. Specially if $I_f=\{k\}$ we recover the simple
formula for $N_k$ for a single oscillator obtained by Greenberg ([2],~[3]).
\ecor

Also, if we plug into (46) and (47) the formulas (22) expressing $Y'_{\bf i}$s in terms
of monomials we obtain Zagier or Stanciu type formulas for number
operator.

The transition operators will be considered in the near future.

{\bf Acknowledgments.}\\
This work was supported by the Ministry of Science and Technology of the
Republic of Croatia under contract No. 0098003 and 037009.
\newpage

 \begin{center}
REFERENCES
 \end{center}

\begin{flushleft}
  [1] Bardek V.,Meljanac S.,Perica A.: Generalized statistics and
 dynamics in curved spacetime.Phys.Lett.B{\bf 338}, 20-22 (1994);\\ {}
  [2] Greenberg O.W.: Example of infinite statistics. Phys.Rev.Lett.{\bf 64}
, 705-708    (1990);\\{}
  [3] Greenberg O.W.: Particles with small violations of Fermi or Bose
statistics.
 Phys.Rev.D{\bf 43}, 4111-4120 (1991);\\{}
  [4] Greenberg O.W.: Interactions of particles having small violations of statistics.
Physica A{\bf 180}, 419-427 (1992);\\{}
  [5] Mohapatra R.N.: Infinite statistics and a possible small violation of the
 Pauli principle. Phys.Lett.{\bf 242B}, 407-410 (1990);\\{}
  [6] M\o ller J.S.: Second quantization in a quon algebra. J.Phys.A:Math.Gen.
{\bf
 26}, 4643-4562 (1993);\\{}
  [7] Meljanac S.,Perica A.: Generalized quon statistics. Mod.Phys.Lett.A{\bf 9}
, 3293-3299 (1994);\\ {}
  [8] Meljanac S.,Perica A.: Number operators in a general
 quon algebra. J.Phys.A:Math.Gen.{\bf 27}, 4737-4744 (1994);\\ {}
 [9] Meljanac S.,Svrtan D.: Study of Gram matrices in Fock representation of
 multiparametric canonical commutation relations, extended Zagier's conjecture,
hyperplane arrangements and quantum groups. Math.Commun.{\bf 1}, 1-24 (1996);\\ {}
 [10] Meljanac S.,Svrtan D. Determinants and inversion of Gram matrices
in Fock representation of $\{ q_{kl}\} $-canonical commutation relations
and applications to hyperplane arrangements and quantum groups. Proof of
an extension of Zagier's conjecture. Preprint RBI-TH-5/Nov.1995;\\{}
 [11] Stanciu S.: The Energy Operator for Infinite Statistics. Commun.Math,Phys.
 {\bf 147}, 211-216 (1992);\\ {}
 [12] Varchenko A.: Bilinear Form of Real Configuration of Hyperplanes. Advances
  in Mathematics {\bf 97}, 110-144 (1993);\\ {}
 [13] Werner R.F.: The free quon gas suffer Gibbs' paradox. Phys.Rev.D{\bf 48},
    2929-2934 (1993);  \\ {}
 [14] Wu Z., Yu T.: Construction of Bose and Fermi operators in terms of
    $q=0$ quon operators. Phys.Lett.A{\bf 179}, 266-270 (1993);\\{}
 [15] Zagier D.: Realizability of a model in infinite statistics, Commun.Math.Ph
 ys.{\bf 147}, 199-210 (1992); \\ {}
 [16] Bo\v zejko M. and Speicher R.: Completely positive maps on Coxeter groups,
 deformed commutation relations, and operator spaces.
 Math.Ann. {\bf 300}, 97-120 (1994); \\ {}
 [17] J\o rgensen P.E.T., Schmitt L.M., Werner R.F.: Positive representations
 of general commutation relations allowing Wick ordering. J.-Funct.-Anal.
{\bf 134}, 33-99 (1995); \\ {}
\end{flushleft}

\end{document}